# Design and Implementation of Data Acquisition and Analysis System for Programming Debugging Process Based On VS Code Plug-In

Boyang Liu
*Operation Department*
*ONUS Global Fulfillment Solutions*
7419 Nelson Rd unit 130, Richmond, BC V6W 1G3, British Columbia Province, Canada
p7908686@gmail.com

*Abstract*—In order to meet the needs of students' programming debugging ability training, this paper designs and implements a data acquisition and analysis system for programming debugging process based on VS Code plug-in, which aims to solve the limitation of traditional assessment methods that are difficult to fully evaluate students' debugging ability. The system supports a variety of programming languages, integrates debugging tasks and data acquisition functions, captures students' debugging behavior in the local editor in real time, and uploads the data to the platform database to realize the whole process monitoring and feedback, provides accurate debugging guidance for teachers, and improves the teaching effect. In terms of data analysis, the system proposed a debugging behavior analysis model based on abstract syntax tree, combined with node annotation, sequence recognition and cluster analysis and other technologies, to automatically track the context of students' debugging process and accurately identify key features in the debugging path. Through this tool, the system realizes the intelligent identification and labeling of the debugging direction and behavior pattern, and improves the refinement level of debugging data analysis. In this research system, a complex debugging scenario of multi-file and multi-task is introduced into the debugging problem design, which optimizes the multi-dimensional capturing ability of debugging data and lays a foundation for accurate debugging behavior analysis. Through several practical teaching tests, the feasibility and stability of the system are verified, which proves that it can effectively support procedural evaluation in programming debugging teaching, and provides a new direction for debugging behavior analysis research.

*Keywords—programming debugging ability, data acquisition and analysis system, VS Code plug-in, abstract syntax tree, debugging behavior analysis*

## I. Introduction

With the development of computer science education, programming course has gradually become the core of computer professional education. Since programming courses involve the combination of complex knowledge system and practical skills, the learning process is always more challenging, especially the programming debugging process, which requires students not only to master grammar and programming logic, but also to have the ability to solve practical problems. The current teaching mode mainly uses written test and computer operation as evaluation methods. This results-oriented evaluation method is difficult to fully reflect the actual ability of students in the process of programming and debugging, especially the process evaluation is relatively scarce, and it is difficult to effectively measure students' programming thinking and debugging skills.

In actual teaching, the training of programming debugging ability usually relies on a large number of experimental operations, but this method has problems of low time efficiency and high repeatability, and students are prone to repeated frustration when they encounter difficulties, and their learning experience is affected. In addition, the individual differences of the debugging process are large, and it is difficult for teachers to provide personalized guidance for each student in large-scale teaching, which further aggravates the consumption of teaching resources. Even if students pass the theory exam and programming test, many still have difficulty completing tasks independently in a real development environment, showing that they "understand the grammar but are not familiar with programming practices".

In order to deal with this series of problems, domestic and foreign researches put forward a variety of measurement and analysis methods based on programming process, trying to improve students' programming debugging ability through data collection and analysis. The current methods are mainly divided into two categories: one is the physiological and psychological measurement methods based on cognitive psychology, such as analyzing the cognitive state of programmers through brain waves or eye tracking technology; The other is to collect the behavior data generated in the programming process through software engineering and human-computer interaction technology, and then analyze the programmer's activity pattern. However, most of these methods are limited to single dimension analysis, lack of multi-dimensional comprehensive measurement of the whole process of programming debugging, and their application in practical teaching is limited.

Aiming at the above problems, this paper designs and implements a data acquisition and analysis system for programming debugging process based on VS Code plug-in based on actual teaching requirements. By integrating a variety of functional modules, the system can collect multi-dimensional data in real time during students' programming and debugging, and use advanced analysis algorithms to process and feedback these data to help students adjust programming ideas in time and improve debugging efficiency. Through the application of this system, teachers can also have a more detailed understanding of students' programming behavior, and then provide more targeted teaching guidance, and finally improve the effect and efficiency of programming teaching.

By constructing a set of perfect data acquisition and analysis tools for programming debugging process, the research in this paper not only provides a means of procedural evaluation for programming teaching, but also provides strong support for the deep understanding and analysis of programming debugging behavior, which has important teaching and practice

## II. RELEVANT RESEARCH

When designing and implementing the data acquisition and analysis system for programming debugging process based on VS Code plug-in, we consulted relevant research results to improve the functionality and performance of the system. Here are some of the key findings and their implications for system design.

J. Warrell proposed a formal probabilistic programming metalanguage [1] that combines cubic type theory and dependent metagraphs to derive synthetic descriptions of type systems through path equivalence. This method provides theoretical support for the system integration of different programming language features, and suggests that we should pay attention to the systematic integration of debugging functions in order to improve the overall efficiency.

Historically, PL/I languages have attempted to combine the strengths of FORTRAN, ALGOL, and COBOL, and although the goal has not been fully realized, this attempt has demonstrated the rationality of integrating the characteristics of different programming languages [2]. This experience shows that when designing VS Code plug-ins, we should integrate multiple debugging functions to improve the compatibility and integration of the system.

A. Neale studied the integration of Buzz programming language with the Pi-puck robot platform [3] and evaluated the effects of packet loss and sensor noise on neighbor perception.

M. Zambrano designed an online course on algorithms and C/C++ programming that uses OOHDM and CROA methods and was tested on SELI and Google Classroom platforms [4]. After analyzing the test results, the researchers found that these methods can effectively organize complex technical educational materials, can significantly improve students' programming skills and get high ratings.

D. Ozdemir proposed a distance education participation detection model based on facial recognition algorithm, which was developed on the Visual Studio.Net platform and carried out image processing through EmguCV library [5]. Through the analysis of the test results of the model, it is found that the accuracy rate is more than 80%. K. Weiss 'CPG analysis platform supports fuzzy parsing and language-level semantic capture by transforming source code into a programming language-independent representation [6]. The results of this study show that the introduction of identity technology in the VS Code plug-in can effectively improve the engagement and data accuracy of the debugging process to improve debugging results.

The UDE 2023 version of PLS introduces several new features for embedded software debugging and runtime analysis, including global time benchmarks and task status displays [7]. The introduction of these functions is important for the accurate monitoring and optimization of system performance, and they can help improve the efficiency and accuracy of the debugging process, thereby optimizing system performance. T. Kano studied the application of eye tracker in program debugging and provided a quantitative analysis method to evaluate debugging behavior by analyzing fixation features and cognitive processes[8].

E. Kamburjan proposed a new approach combining programming language and semantic technology, creating a deeply integrated semantic reflection mechanism by directly mapping program state to RDF knowledge graph [9]. This innovative method provides a new way to realize the close combination of program state and semantics, and is suitable for debugging and verification functions in VS Code plug-in, thus improving the intelligence level of the system.

The Virtual Educational Robot system (AR Bot) developed by FCO Yang uses augmented reality technology to provide 3D visual feedback and automatic scoring, which significantly improves students' computational thinking ability [10]. The method of automatically detecting JTAG debugger/emulator invented by RG Krishna recognizes debugging mode by sending reset signal and detecting programming signal [11]. This technology provides an important reference for automatic detection and debugging in VS Code plug-in, which can improve the automation level of the system and enhance the accuracy and efficiency of the debugging process.

These research results provide rich theoretical and technical support for the design of data acquisition and analysis system in programming debugging process based on VS Code plug-in. By integrating advanced technologies and methods, the debugging process can be optimized to improve the functionality, accuracy and user experience of the system.

## III. PROGRAMMING AND DEBUGGING DATA ACQUISITION AND ANALYSIS TOOL DESIGN

### A Programming and debugging teaching demand analysis and topic design ideas

In the process of programming and debugging teaching, students' learning behavior is more complicated than the traditional learning mode in the classroom. Different from the traditional teaching mode in which teachers are mainly responsible for teaching and students mainly memorize, programming teaching not only requires teachers to pass on basic grammar and related knowledge, but also requires teachers to pay special attention to training students' programming debugging ability in the teaching process. In the whole process of learning programming, students not only need to be familiar with specific knowledge points, but also must combine the learned knowledge with the actual programming debugging process through independent learning and cooperation with others, so as to solve practical problems or complete programming projects. Due to the complexity of this learning mode, it is difficult for teachers to rely on traditional teaching methods to fully grasp the learning state of students in the process of programming and debugging, especially in the face of students' debugging behavior and thinking mode, teachers often fail to clearly understand the specific problems encountered by students or the operation path. In order to better support teaching, it is necessary to design a set of data acquisition and analysis system that can capture and analyze students' programming and debugging behavior in real time, so as to help teachers get more detailed feedback and provide students with more accurate teaching guidance.

At present, many programming debugging teaching tools can only collect students' debugging data through online editors, and the limitation of such tools is that they cannot capture the actual operational data of students' programming debugging while using local development tools. Due to this limitation, it is difficult for teachers to obtain the debugging problems and behavioral data that students actually encounter in the local development environment, which makes teachers unable to accurately grasp the real learning process of students. Therefore, in response to this teaching demand, the system is designed based on VS Code plug-in form, through the integration of a series of special data collection functions, under the premise of not affecting students' habits, it can collect students' programming debugging data in various teaching scenarios. By capturing key information such as snapshots of code runtime, code modification records, error logs, and debug operations, the system can comprehensively record every step of the student's programming process. This data not only contains the basic information of compilation and running, but also covers the common fallback operations, debugging steps, and ways to solve problems in the debugging process. Through these detailed data, teachers can analyze students' debugging behavior more deeply, so as to provide students with more targeted feedback, help students timely correct mistakes and optimize debugging ideas.

Topic design in programming teaching plays an important role in cultivating students' debugging ability. Based on the comprehensive data collection function provided by the system, the topic design should not only cover the required knowledge points, but also focus on guiding students to strengthen their programming thinking through practical debugging and problem solving. With the help of this system, teachers can design step-by-step programming debugging problems according to students' learning situation, so that students can gradually transition from dealing with simple grammar errors to dealing with more complex logic problems, so as to help them build effective debugging methods and thinking models. The system also has the function of providing personalized feedback according to the real-time debugging data of students, and can dynamically adjust the difficulty of the problem and provide relevant prompt information. For example, when students repeatedly encounter problems with a specific type of error, the system can provide them with specialized tips based on the captured data, helping students better understand the problem and finally solve the problem, thus further improving the personalized, targeted and overall efficiency of teaching.

*B   3.2 Overall tool structure and module function design*

Although the current programming debugging tools have certain functions in data collection, their application in the actual teaching scene is still insufficient, especially in the direct feedback programming debugging process. This limits its role in assisting teaching. In order to enhance this capability, we plan to design a new set of analysis tools on the "Nirvana Road" learning platform, focusing on the addition of programming debugging process feedback module. This module will be based on abstract syntax tree to realize the visual analysis of debugging process code snapshot.

The platform will share a server with data acquisition tools to integrate programming debug records from different users. Therefore, the platform needs a unified login system to verify user identity and ensure the accuracy of information. At the same time, the platform will use the small program code snapshot provided by the collection tool stored in the programming debugging process of the server as the input of the feedback module. In the folder of the applet, the view layer code (WXML file), logic layer code (JS file), and style layer code (WXSS file) will be handled independently, each playing its own function and using its characteristics for analysis.

In the specific analysis process, the platform will use different methods and third-party tools to conduct targeted analysis according to the file structure and content characteristics of the wechat mini program, combined with the time series characteristics and clustering characteristics of the user debugging process. Therefore, we will develop special analysis tools for the programming debugging process, aiming to improve the user's understanding of the debugging process, and carry out qualitative and quantitative statistical analysis. Analysis tool design falls into two categories: general purpose tools and statistical analysis tools. Common tools include JS source code translation abstract syntax tree tool, time-based debugging process visualization tool and debugging process visualization tool based on the same file. Statistical analysis tools include wechat API statistical analysis tool based on abstract syntax tree, JS file flow chart differentiation detection tool, abstract syntax tree and source code conversion tool, debugging process behavior identification tool and debugging direction annotation tool.

In the implementation of the tool, abstract syntax tree is an abstract representation of the structure of the source code language, and shows the syntax elements of the programming language through the tree structure. Although it does not cover all the syntax details, it condenses the core components of the syntax structure, provides a simple operation interface, and facilitates the in-depth exploration of the programming debugging process. With the help of abstract syntax trees, we can build clear code descriptions that facilitate subsequent modifications, transformations, and annotations of the source code.

In order to improve the analysis ability of programming debugging tools, we will combine specific professional data tables to provide more detailed performance comparison and optimization basis. Table 1 shows the loading speed of different parsers under various programming frameworks. According to these data at the front end, the platform will obtain the code snapshot of the programming debugging process collected by the plug-in from the back-end database and screen out the JS file. For JS files containing syntax errors, the system will not be able to generate an abstract syntax tree, and the front end will catch these errors using a try-catch statement to get an error-free snapshot of the JS file code. Finally, these code snapshots are translated and corresponding abstract syntax trees are generated to complete the comprehensive analysis of the programming debugging process.

TABLE 1. COMPARISON OF PARSER LOADING SPEEDS

| Parser | Frame 1 Loading speed (ms) | Frame 2 Loading speed (ms) | Frame 3 Loading speed (ms) | Frame 4 Loading speed (ms) | Frame 5 Loading speed (ms) | Average loading speed (ms) |
|---|---|---|---|---|---|---|
| Esprema | 219.4 | 247.2 | 231.5 | 207.0 | 290.4 | 239.1 |
| UglifyJS | 182.8 | 234.2 | 235.5 | 190.7 | 236.4 | 215.3 |
| Traceur | 216.7 | 240.8 | 213.4 | 200.9 | 260.6 | 218.5 |
| Acorn | 182.1 | 228.5 | 221.7 | 186.2 | 239.8 | 211.6 |
| Esprima | 197.3 | 252.6 | 236.4 | 208.5 | 271.0 | 233.2 |

In the functional design of developer tools, the implementation of interactive functions aims to improve the user's programming experience and solve common problems. The design of the reset workspace function is based on an analysis of the actual needs of students when using developer tools. Through questionnaires and test feedback, we found that students often had to start from scratch due to the wrong direction of debugging in the process of debugging, but because they could not accurately trace the debugging steps and behaviors, they often had to return the code by memory or continue to debug on the current code basis. Therefore, the function of resetting the workspace allows users to return to the initial state of the code with one click, which can significantly improve the efficiency of programming debugging. The implementation involves storing the initial code snapshot in the workspace after the plug-in has taken it. When the user chooses to reset the workspace, the plug-in calls the initial code snapshot stored, and restores the code to the original state through the file overwrite function, thus achieving the debugging rollback.

Breakpoint continue-answer function is designed to solve the problem of unexpected interruption that students may encounter in the process of programming debugging. Through the investigation, it was found that after students entered the ID of the question, sometimes they accidentally closed the developer tool without formally submitting it. To solve this problem, the server automatically ends the Session and retains the created session object after the session times out, so that the user can continue the progress of the last debugging when entering the debugging state again. The server will determine the status of the last session according to the topic ID, and return the latest code snapshot after detecting the session timeout to realize the resumable function.

The help and question solving function is designed to take into account the different needs of students and teaching assistants. Students only have the right to help, while teaching assistants have the dual right to help and solve questions. Through actual research, we learned that students want to be able to share the debugging process and receive direct guidance, so the help function is set up under the command tree at all levels of the student and teaching assistant status to record and solve programming debugging problems. The help function is triggered by clicking the "help" command. After students fill in the help form and submit the code snapshot, the teaching assistant can solve the question according to the problem ID. In the question and answer module of the Nirvana Path learning platform, students can view the list and details of all the questions asked for help, and the teaching assistants' problem-solving process and code snapshots are also publicly displayed for students to discuss and learn.

The troubleshooting function is triggered by the "troubleshooting" command. The process includes filling in the problem ID, debugging the code locally, viewing the help form, and submitting the code snapshot. During the troubleshooting process, the server records every save and compile operation of the TA and analyzes it at the end of the session to show the difference between the debugging process of the changed file and the snapshot of the original code. This allows students to get targeted guidance and improve their programming debugging skills.

For problem design, emphasis is placed on the design of parametric errors rather than functional errors during programming debugging. The debugging topic aims to reduce students' editing freedom and provide clear debugging direction. The design of the questions takes into account the needs of actual teaching, mainly involving the types of errors in a single file, while taking into account the different languages and types of errors in the wechat mini program, to ensure that the questions can effectively measure the debugging ability of students.

The practice questions are designed for the teaching scene where students write questions for each other, and are edited and published directly by students without the authorization of teaching assistants. The role of the teaching assistant is to guide the students to follow the principles of purpose, variety and hierarchy in the design of the topic. The acceptance questions correspond to the homework assigned by the teacher, and the rigor and difficulty gradient of the questions should be fully considered in the design, and strict testing should be conducted before the official release to ensure that the program has no other problems except design errors. Through the functional testing and error testing process, we ensure that the designed questions can effectively test the debugging ability of students, and that the questions are fully tested and verified before publication.

IV. IMPLEMENTATION AND APPLICATION OF DATA ACQUISITION AND ANALYSIS TOOLS

A *Programming process data acquisition and function realization*

When describing the platform's development environment and code structure in detail, we first need to fully understand its components and what they do. The platform's development environment includes a range of tools and technologies, each of which plays an important role in ensuring a smooth development process. The main development tools include the Code editing tool Visual Studio Code and front-end build tools such as Webpack and Rollup, which are responsible for the packaging and compression of the code, making the deployment and update of the front-end code efficient.

In terms of the realization of plug-in related functions, the platform's plug-in functions include three main parts: help and doubt solving, bug challenge and free debugging mode. The Help and Question solving feature allows users to ask for help by filling out question forms, while the bug Challenge feature includes sub-features such as creating rank, rank mode, and training mode, which are designed to help users detect and solve problems in their code. Free debugging mode provides the ability to pull down other people's code, get code from the code base, and use local code, which greatly enhances the flexibility and convenience of debugging. Overall, through the design and implementation of these functions, the platform not only improves the development efficiency, but also enhances the usability and user experience of the system.

In terms of resource packaging optimization, GitLab platform mainly includes two aspects: resource compression and resource segmentation. Resource compression reduces the size of the packaged project file by removing comments, redundant symbols, Spaces, and invalid functions from the code. Resource segmentation Divides resources into modules and loads them on demand.

*B Tool Performance testing and analysis*

After the completion of system development, software testing is a crucial part. This chapter will introduce the test scheme and results of the data acquisition and analysis tool for programming process in detail.

The design of the test case is based on the specific design of the functional module of the tool. On this basis, we list the core functional test cases of the tool, which are detailed in the following table. In testing the login plug-in, we verify that the plug-in can successfully connect to the unified login platform. The operation process includes clicking the "Login plug-in" button, the system should jump to the unified login platform, and ask the user to enter the account and password. After successful login, the system will generate a token, the user needs to copy and paste it into the developer tool input box, click the "login" button to complete the login process.

When testing the "bug Challenge" feature, we checked that the plug-in was able to correctly handle user-created challenges when system service was normal. The test steps include clicking on the plugin command, selecting "bug Challenge" and creating a row. We verify that the plugin can enter debugging state after completing functions such as unquestioning and creating rows, and check that the command tree is updated. After debugging, the system should clean the workspace and update the relevant information. In testing the debugging process visualization tool, we focused on whether the system can display a snapshot of the debugging process according to the time change, and provide an accurate view of the debugging progress and results. The tests include checking the visualizations of different file types and code flows, as well as the handling of abstract syntax trees.

In the course of comprehensive testing and analysis of the tool's performance, we evaluated its performance from multiple perspectives, including page loading performance, interface response performance, and statistical analysis of user data.

Tests of the tool's page loading performance show that the first screen loading time of the tool is about 5000 milliseconds without optimization. This long load time is mainly due to the large number of external static resources introduced in plug-ins and single-page applications. To solve this problem, we optimized the packaging process of the platform and introduced a resource caching mechanism. After the optimization was implemented, the resource load time of the platform was significantly reduced to 1700 ms. In this way, we reduced the first screen load time by 66%, which effectively improved the user experience.

In terms of interface response performance, we conducted a stress test to evaluate the stability and reliability of the tool's back-end interface under high load conditions. During the test, 1000 requests were sent to the server interface per second, and the test results showed that the client successfully received all the responses from the backend interface. This result proves the stability of the interface under high concurrency conditions and verifies its ability to meet the practical teaching needs.

For user data analysis, we conducted in-depth statistical analysis of the tool's collection and annotation functions. For example, in the OCR recognition translation function test, we collected 473 user debugging records, as shown in Table 2. These records reveal that complex program errors typically require more debugging times than simple parameter, attribute, and syntax errors, as shown in Figures 1, 2, and 3. In addition, the difficulty of debugging problems has a significant impact on the number of debugging times, with more difficult problems often requiring more debugging times, while less difficult problems have relatively fewer debugging times. This difference shows that there is a significant correlation between the number of debugs and the difficulty of the problem.

TABLE 2. PERFORMANCE TEST AND USER DATA STATISTICS

| Question type | Total number of users | Number of debugs | The number of debugs per session | The number of times each question is debugged | The average number of debugs | Total number of debugs | Completions |
|---|---|---|---|---|---|---|---|
| OCR recognition | 473 | 1186 | 1.5 | 16 | 5.1 | 5150 | 1940 |
| Translation | 404 | 828 | 1.5 | 12 | 5.9 | 5580 | 1595 |
| Board games | 99 | 730 | 1.5 | 14 | 6.4 | 1215 | 664 |

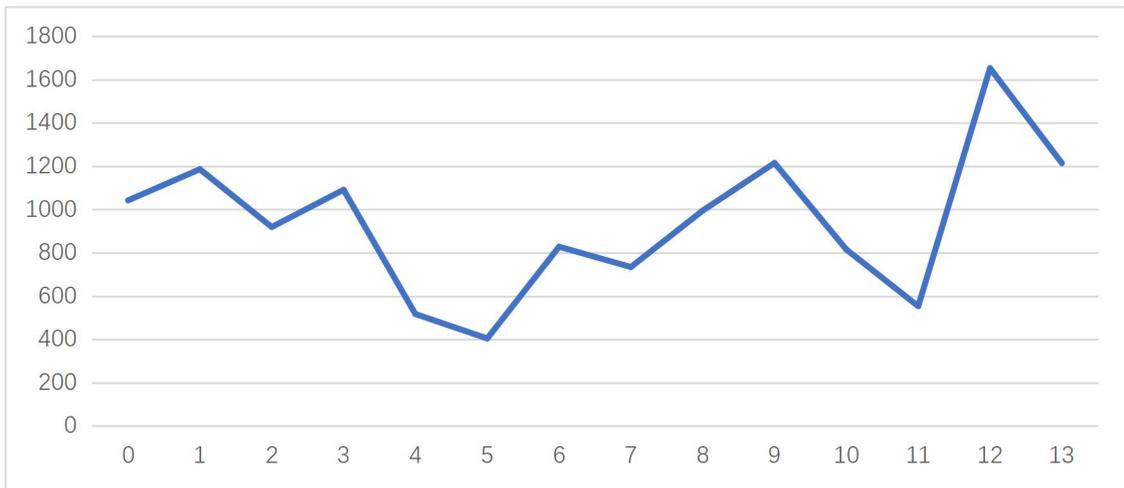

Figure 1. Debugging times of each OCR question

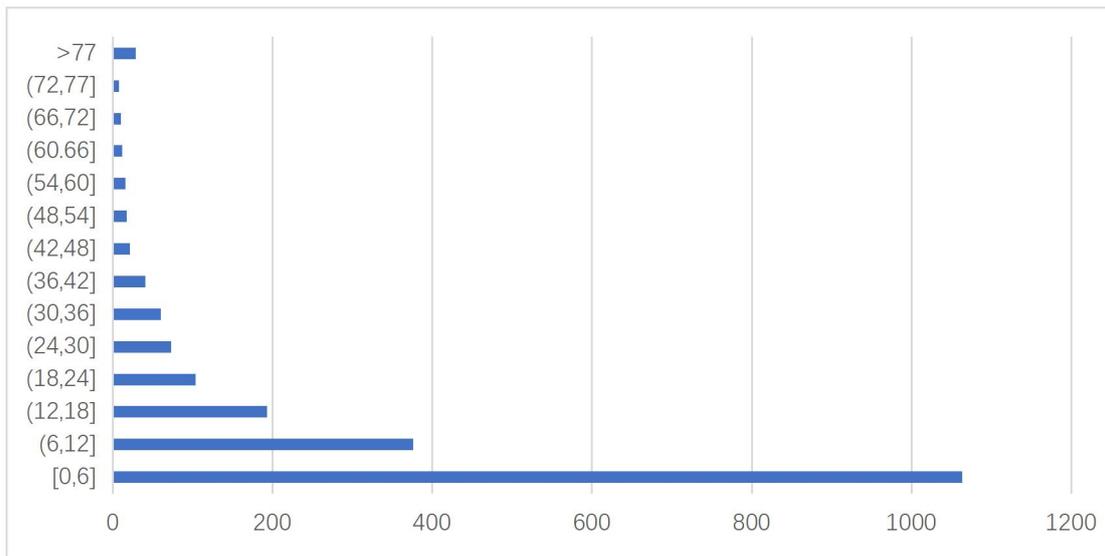

Figure 2. Debugging times of OCR identification and translation problems

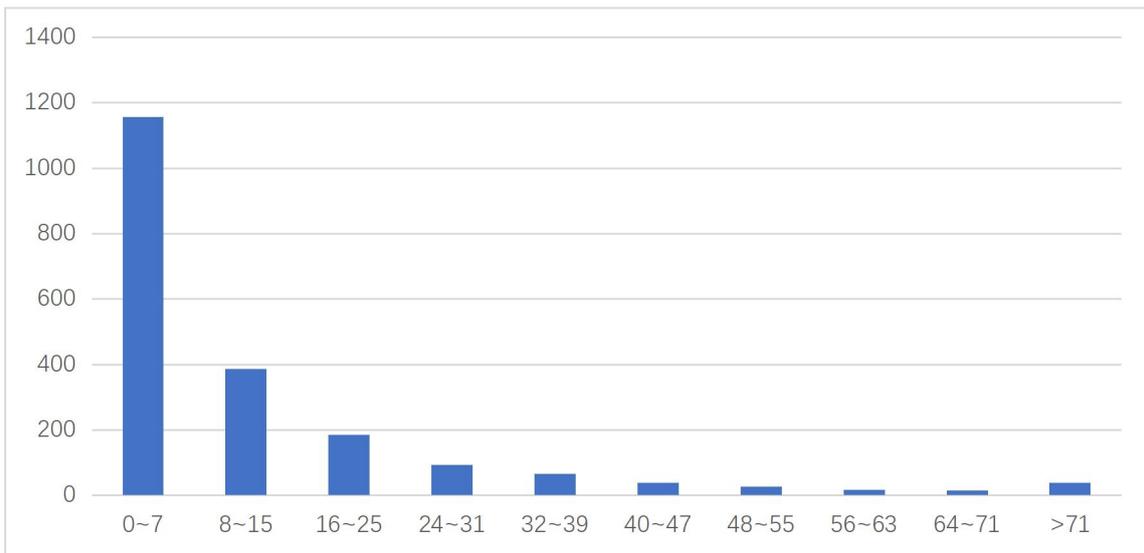

Figure 3. OCR recognition and translation sessions and debugging times

V. CONCLUSION AND PROSPECT

This research successfully developed a data acquisition and analysis tool focusing on programming and debugging process, which captured debugging data in real time through VS Code plug-in and uploaded it to the platform, thus providing accurate feedback for teachers and students. The tool combines multi-file and multi-error debugging problem design to comprehensively analyze students' debugging process and support multi-dimensional data capture and analysis. Although the current version has demonstrated stability and effectiveness, there is still room for improvement in data acquisition and error handling capabilities, such as expanding the dimensions of data acquisition to cover mouse and keyboard information and eye movements, and optimizing the automatic labeling of functional errors. Future work will focus on further enhancing the comprehensiveness and compatibility of the tools to better support programming debugging teaching, enhance student learning, and meet changing educational needs through continuous optimization.

## REFERENCES

[1] Warrell J , Potapov A , Vandervorst A ,et al.A Meta-Probabilistic-Programming Language forBisimulation ofProbabilistic andNon-Well-Founded Type Systems[C]//International Conference on Artificial General Intelligence.Springer, Cham, 2023.DOI:10.1007/978-3-031-19907-3_42.

[2] Meil D .Lessons from PL/I: A Most Ambitious Programming Language[J].Communications of the ACM, 2023, 66(11):6-7.

[3] Neale A , Millard A G .Integration andRobustness Analysis oftheBuzz Swarm Programming Language withthePi-puck Robot Platform[C]//Annual Conference Towards Autonomous Robotic Systems.Springer, Cham, 2022.DOI:10.1007/978-3-031-15908-4_18.

[4] Zambrano M , Villacis C , Alvarado D ,et al.Active Learning of Programming as a Complex Technology Applying Problem Solving, Programming Case Study and OnlineGDB Compiler[C]//2021 10th International Conference on Educational and Information Technology (ICEIT).2021.DOI:10.1109/ICEIT51700.2021.9375611.

[5] Ozdemir D , Uur M E .A Model Proposal on The Determination of Student Attendance in Distance Education with Face Recognition Technology[J].Turkish Online Journal of Distance Education, 2021, 22(1):19-32.DOI:10.17718/tojde.849872.

[6] Weiss K , Banse C .A Language-Independent Analysis Platform for Source Code[J]. 2022.DOI:10.48550/arXiv.2203.08424.

[7] Morling R .PLS' UDE 2023 simplifies debugging and trace analysis for high-end MCUs[J].Design Solutions, 2023(Feb.).

[8] Kano T , Sakagami R , Akakura T .Modeling of cognitive processes based on gaze transition during programming debugging[J].IEEE, 2021.DOI:10.1109/LifeTech52111.2021.9391940.

[9] Kamburjan E , Klungre V N , Schlatte R ,et al.Programming and Debugging with Semantically Lifted States[J].Springer, Cham, 2021.DOI:10.1007/978-3-030-77385-4_8.

[10] Yang F C O , Lai H M , Wang Y W .Effect of augmented reality-based virtual educational robotics on programming students' enjoyment of learning, computational thinking skills, and academic achievement[J].Computers & Education, 2023, 195:104721-.DOI:10.1016/j.compedu.2022.104721.

[11] Krishna R G , Rao S M .AUTO DETECTION OF JTAG DEBUGGERS/EMULATORS:EP20190176860[P].EP3575807B1[2024-09-17].